\documentclass[fleqn,usenatbib]{mnras} 

\usepackage[T1]{fontenc}
\usepackage{ae,aecompl}
\usepackage{graphicx}	
\usepackage{amsmath}	
\usepackage{amssymb}	

\newcommand{\msun} {${\cal M}_\odot$}

\title[Formation of close binaries by migration]{Formation of close binaries by disc fragmentation and migration, and its  statistical modeling}

\author[Tokovinin \& Moe]{Andrei Tokovinin,$^{1}$\thanks{E-mail: atokovinin@ctio.noao.edu}
and Maxwell Moe$^{2}$ 
\\
$^{1}$Cerro Tololo Inter-American Observatory, Casilla 603, La Serena, Chile \\
$^{2}$Steward Observatory, University of Arizona, 933 N. Cherry Ave., Tucson, AZ 85721, USA
}

\date{Accepted XXX. Received YYY; in original form ZZZ}
\pubyear{2019}

\begin{document}
\label{firstpage}
\pagerange{\pageref{firstpage}--\pageref{lastpage}}
\maketitle

\begin{abstract}
Joint statistics of periods and  mass ratios of close binaries and its dependence on primary  mass can  be explained  by assuming that seed binary  companions are  formed by  disc fragmentation at random intervals during assemblage of stellar mass and migrate inwards as they accrete from the circumbinary disk.  A toy model based on simple prescriptions for the companion  growth  and  migration  reproduces such aspects  of  close solar-mass binaries as the distribution  of binary periods $P$,  the brown  dwarf  desert at  short  $P$, the  nearly uniform distribution of mass ratios, and a population  of equal-mass binaries  (twins) that decreases linearly in frequency with log\,$P$. For massive  stars, the model predicts a large fraction  of early mergers, a distribution of $\log  P$ with  a  negative   slope,  and a mass-ratio distribution that is also uniform but with a substantially reduced twin fraction. By treating disc fragmentation as a stochastic process, we also reproduce the observed properties of compact triples. Success  of our  toy model suggests  that most  close binaries and compact triples indeed formed by disc fragmentation followed by accretion-driven inward migration.
\end{abstract} 

\section{Introduction}
\label{sec:intro}

It is generally accepted that  most binary stars form by fragmentation of  proto-stellar cores  or  circumstellar discs \citep{Bate1995,Kroupa1995,Bate2002,Tohline2002,Kratter2006,Clarke2009,Offner2010,Kratter2016,Moe2017,Moe2019}   In  both cases,  the initial separations  cannot be  less than $\sim$10~au because of the opacity limit to fragmentation \citep{Boss1986,Bate1998}. The first hydrostatic cores form with radii of a few au, and fragmentation during the secondary collapse phase is unlikely \citep{Bate1998, Bate2011}. At such close separations, the accreting gas is generally too hot to become self-gravitating and fragment. Even if the opacities were smaller, e.g., at lower metallicities,  any small density perturbations in the gas would quickly redistribute because the Keplerian orbital periods are shorter than the cooling timescales \citep{Moe2019}. The observed population of close binaries with $a$~$\lesssim$~10~au must have initially fragmented at wider separations and subsequently migrated inward \citep{Bate2002}.

In very dense environments like globular clusters, close binaries may form via tidal capture, disk capture, and/or N-body dynamical interactions \citep{Press1977,Murray1991,Hurley2007,Sollima2008}. Close field binaries, which formed in low density environments, require an alternative explanation for inward migration, such as hydrodynamical forces and torques in a circumbinary disk,  orbital decay from protostellar accretion, and gravitational interactions in triple stars \citep{Artymowicz1983, Artymowicz1991,Bate1995,Bate1997,KCTF,Reipurth2001b}. The majority of solar-type binaries with semimajor axis $a$~=~0.1\,-\,10~au do not have additional companions \citep{Tok2006,Tok2014}, and so most close binaries migrated without the assistance of triple star interactions. Meanwhile, a significant majority of very close solar-type binaries with $P$~$\lesssim$~10~days ($a$~$\lesssim$~0.1~au) have tertiary companions \citep{Tok2006}. \citet{KCTF} suggested Kozai-Lidov interactions in misaligned triple stars, coupled to tidal friction, produce such very close pairs.  Assuming all triple stars have random orientations, \citet{Fabrycky2007} and \citet{Naoz2014} simulated Kozai-Lidov interactions for 10~Gyr and matched the observed properties of very close binaries. 

However, \citet{Moe2018} showed that this mechanism can generate only  a  small fraction  of very close  binaries due to two main effects. First, the majority of compact triples, especially those with outer tertiaries $a_{\rm out}$ $<$ 10 au (below the opacity limit),  actually have quasi-planar architectures \citep{Borkovits2016,Tok2017}. Not only does this severely limit the fraction of triples that undergo Kozai-Lidov oscillations, but points to  the  dominant role of disk fragmentation and migration in the formation of close binaries and compact triples (see also \citealt{Tobin2016}).  Second, \citet{Moe2018} noted that the very close binary fractions of solar-mass pre-main-sequence (pre-MS) and field binaries are nearly identical \citep{Mathieu1994,Melo2003}. In particular, \citet{Kounkel2019} recently demonstrated that class II/III T Tauri stars exhibit the same binary fraction and period distribution below $P$~$<$~10$^4$~days ($a$ $<$ 10 au) as their field counterparts, with at most a $\approx$30\% deficit at the shortest of periods $P$~$<$~5~days. Close binaries with $a$~$<$~10~au and the majority of very close binaries with $a$ $<$~0.1~au must have migrated during the embedded Class 0/I phase (age $\tau$ ~$\lesssim$~2~Myr) while there was still dissipative gas in the surrounding disk and envelope.

The physics of core and disc fragmentation set the initial masses of binary components. The seed components subsequently grow into stars by accretion.  Components of wide binaries that form via core fragmentation tend to accrete from their respective gas reservoirs and form their own circumstellar disks \citep{Bate1995,White2001,Offner2010,Bate2014}. Meanwhile, closer proto-binaries that derive from disk fragmentation clear out an inner cavity and subsequently accrete from a circumbinary disk \citep{Artymowicz1983, Artymowicz1991,Kratter2006,Clarke2009}. Accretion of gas onto a binary modifies its orbital separation, eccentricity, mass ratio, and the orientations of stellar rotation axes  (spins), a process sometimes referred to as eigen-evolution \citep{Kroupa1995}.  Therefore,  statistics of close  binaries are determined  by their  accretion-driven migration.  Here we  propose a simple mathematical model of disk fragmentation, accretion, and migration that matches qualitatively the  joint  statistics  of  periods  and mass  ratios  of  real  close binaries. 

Some close binaries may also derive from dissipative capture of seed protostars formed by core fragmentation as they initially fell to the common centre of
gravity via dynamical friction with the surrounding gas, and then subsequently accreted and migrated further inward within a circumbinary disk \citep{Offner2010,Bate2019}, see also (Lee et al., submitted). This channel might be more relevant for close M-dwarf binaries where disk fragmentation is less likely \citep{Kratter2010,Offner2010}. For solar-type systems, however, there is strong observational evidence that a significant majority of companions within $a$~$<$~10~au derived from disk fragmentation and migration. As already indicated, $\approx$90\% of compact solar-type triples with $a_{\rm out}$~$<$~10~au have small mutual inclinations $i_{\rm mut}$ $<$ 40$^{\circ}$ \citep{Borkovits2016}. Perhaps more compelling is the dependence on metallicity: the close binary fraction of solar-type stars ($a$~$<$~10~au) is strongly anti-correlated with metallicity because optically thick disks become cooler and more prone to fragmentation with decreasing metallicity, but the wide binary fraction ($a$ $>$ 200 au) is independent of metallicity because fragmentation of optically thin cores is metallicity invariant \citep{Badenes2018,Moe2019,ElBadry2019}. 
Quantitatively, \citet{Moe2019} measured the binary fraction within $a$~$<$~10 au to decrease by a factor of $\approx$4 across $-$1.0 $<$ [Fe/H] $<$ 0.5, but found the underlying separation distribution across $a$~=~0.1~-~10~au to be insensitive to metallicity. \citet{ElBadry2019} confirmed the emergence of a metallicity dependence within $a$~$<$~200~au, demonstrating the fraction of solar-type primaries with $a$~$\approx$~50~au companions decreases by a factor of $\approx$3 across $-$1.0 $<$ [Fe/H] $<$ 0.5. The observations suggest $\gtrsim$90\%, $\approx$70\%, and $\lesssim$10\% of solar-type binaries with $a$~$<$~10~au, $a$~$\approx$~50~au, and $a$~$>$~200~au, respectively, derived from disk fragmentation. In any case, our model proposed here is applicable as long as the seed binary accretes most of its mass from a common circumbinary gas reservoir. For simplicity, we discuss only close solar-type and early-type binaries with $a$~$<$~10~au that most likely formed via disk fragmentation. 

The  complexity  and inherently stochastic  nature of  binary star formation have severely limited comparison of predictive models of binary  statistics with observations. State-of-the-art hydrodynamical simulations of  a dense collapsing cluster   by   \citet[][and earlier versions]{Bate2019}   nicely   demonstrate   this complexity by generating  a  population  of binaries  and  triples resembling real systems in many  ways.  However, the limited number of objects  formed in  these closed-box simulations  precludes  detailed statistical comparison with  observations, especially for relatively rare but astrophysically important massive OB stars. Moreover, the spatial resolution  of these simulations  is   too  coarse   for producing  close   binaries  with separations less than a few au. Finally, not all stars are born in such a dense cluster; formation of stars  and stellar systems depends on the environment.

Section~\ref{sec:model} discusses formation of binaries by disk fragmentation, their accretion-driven evolution, and
the toy model of these processes developed here. Resulting statistics of simulated populations  of solar-type and B-type binaries are presented in Section~\ref{sec:res}. The paper closes by a short discussion of the results in Section~\ref{sec:disc}.

\section{Toy Model of Binary Formation}
\label{sec:model}

The  toy model  presented here  hides the  complexity  of close-binary formation  behind simple  prescriptions with  random  parameters.  Our intent  is to capture  the essential  aspects of  this process  and to reach a qualitative agreement  with the observed statistics of periods $P$  and  mass  ratios  $q$  of close  binaries.  The  model  does not explicitly consider eccentricity, spin vectors, or mutual inclinations between binary orbits and discs, but instead simply averages across these quantities. We simulate companions that derive from fragmentation, accretion, and migration in the disk, and so ignore companions that form beyond $a$~$\gtrsim$~3,000 au via turbulent fragmentation of molecular cores.   We also simulate triples, for which we model the fragmentation, accretion, and migration of the inner binaries and outer tertiaries independently. We present our baseline model parameters for solar-type and early-B primaries in Table~\ref{tab:par}, and we explore different parameter ranges in our supplementary models (Section~\ref{sec:par}). We  adjust  the parameters of our baseline  model to mimic the real binary  statistics, at least qualitatively (one cannot expect a perfect match from our crude model). Most importantly, by simulating large populations of multiple stars using a model with only a few free parameters, we can directly investigate how certain physical processes affect the properties of close binaries, e.g., how weighting disk fragmentation toward earlier or later times changes the mass-ratio distribution, or how angular momentum transfer from the circumbinary disk to binary components alter the degree of correlation between periods and mass ratios.

\begin{table*}
\center
\caption{Parameters of our baseline model.}
\label{tab:par}
\begin{tabular}{l l c  c} 
\hline
Parameter   & Description & Solar-type & B-type    \\
\hline
$M_{\rm tot,0}$, $M_{\rm tot,1}$  & Primary mass range, \msun & [0.7, 1.3] & [10, 20] \\ 
$K_{\rm step}$ & Number of accretion episodes & 20   & 50 \\
$f_{\rm bin}$ & Average number of companions & 0.3  & 2.0 \\
$f_{\rm m0}$ & Initial primary mass $m_0 = f_{m0} M_{\rm tot}$  & 0.1   & 0.1 \\
$f_{\rm 20}$ & Initial companion mass, fraction of $m_{\rm acc}$  & 0.25 & 0.25 \\
$f_{\rm m2max}$ & Maximum companion growth in one episode & 1.0  & 1.0 \\
$a_0$, $a_1$ & Initial separations, au & [40, 1000] & [40, 3000] \\
$\beta$ & Parametrization of the $q$-dependence & 0.7 & 0.7 \\
$\eta_0$, $\eta_1$ & Range of migration coefficient $\eta$ & [0, 3] & [0, 4] \\
\hline
\end{tabular}
\end{table*}

\subsection{Initial Conditions}
\label{IC}

Each system starts as a single star of seed mass $m_0$ = 0.1$M_{\rm tot}$, where $M_{\rm tot}$ will be the final total mass of the system, e.g., $M_1$ for single stars, $M_1$+$M_2$ for binaries, and $M_1$+$M_2$+$M_3$ for triples.  We explore different primary seed masses of $m_0$ = 0.05$M_{\rm tot}$ and 0.15$M_{\rm tot}$ in models {\tt PrimSeed1} and {\tt PrimSeed2}, respectively. Multiplicity statistics are typically measured from observations across a narrow range of primary masses $M_1$~=[M$_{\rm 1,low}$,M$_{\rm 1,high}$], not system masses. We consider solar-type systems to span $M_1$~=~0.7\,-\,1.3\,\msun\ while early-B primaries have $M_1$~=~10\,-\,20\,\msun. To encompass all possible combinations, we select the total final mass uniformly across $M_{\rm tot}$ = [M$_{\rm 1,low}$,3M$_{\rm 1,high}$]. Some simulated primaries will accrete above $M_1$~$>$~M$_{\rm 1,high}$, depending on the fragmentation and accretion evolution of the system. For each model, we remove such systems with massive primaries and simulate until there are $N_{\rm prim}$ = 10$^3$ (or more) systems with final primary masses across $M_1$~=[M$_{\rm 1,low}$,M$_{\rm 1,high}$]. We keep track of the total number $N_{\rm tot}$ $>$ $N_{\rm prim}$  of systems actually simulated per model.

Protostellar  accretion is  a highly stochastic process. Infalling streams  of gas  are  likely episodic.   Moreover, protostellar  disks experience recurrent thermal instabilities, which are manifested as FU Orionis  outbursts.   \citet{Hartmann1996}  estimated  that  young low-mass protostars undergo $\sim$20 FU Orionis outbursts during their initial $\sim$\,1-2 Myr. Discs tend to fragment when their gas temperatures become too cool and/or their accretion rates becomes too high so as to drive a gravitational instability according to the Toomre Q criterion \citep{Kratter2006,Kratter2008,Machida2009,Tanaka2014,Kratter2016,Moe2019}. Episodic accretion is  therefore essential for disk  fragmentation in two  ways. First,  the accretion rate during  a  burst  is much larger than its average value, promoting instability and fragmentation. Second,  the accretion  energy is  radiated away  between  the bursts, and therefore the  disc temperature remains low before the subsequent burst. The infalling gas may not even settle into a viscous-supported disc but rather fragment almost immediately, as in the simulations by \citet{Goodwin2004}. \citet{Kratter2008}, \citet{Tanaka2014}, and \citet{Moe2019} all found that discs of solar-type stars at solar metallicity are unlikely to fragment if they accrete constantly at their average rate of $\dot{M}$ $\approx$ $10^{-6}$\msun\,yr$^{-1}$. \citet{Moe2019}  emphasized that variability in the accretion rate is required; stochastic excursions up to $\dot{M}$ $\approx$ $10^{-5}$\msun\,yr$^{-1}$, i.e., $\sim$10 times the average accretion rate, were sufficient to fragment the disc (see their Fig. 20). In our toy model, mass accretion occurs in $K_{\rm  step}$ discrete episodes or intervals, with the mass $\Delta m$ = $(M_{\rm tot} - m_0)/K_{\rm   step}$ accreted in each episode. For our baseline model, we adopt $K_{\rm step}$ = 20 and 50 for solar-type and early-B systems, respectively. We explore half and double these values in models {\tt Step1} and {\tt Step2}, respectively. At each accretion episode, a companion can form  with a certain probability (see below). Physical time is not  involved in our model, and by  ``time'' $t$ = $m$/$M_{\rm tot}$ we refer to the fraction of accreted mass.

Class II and even class I T Tauri disks, which have masses $M_{\rm disk}$ $\sim$ 0.01\,$M_1$ and $\sim$ 0.1\,$M_1$, respectively \citep{Sheehan2014,Ansdell2016,Sheehan2017}, are unlikely to fragment due their small disc masses and low accretion rates \citep{Kratter2016}. As discussed in Section \ref{sec:intro}, disc fragmentation and accretion must have occurred in large part within the initial $\tau$~$<$~2~Myr because the close binary properties of T~Tauri stars match the field values \citep{Kounkel2019}.

The  average frequency of companions  formed  during  the  mass  buildup is  determined  by  the free parameter $f_{\rm bin}$.  We emphasize $f_{\rm bin}$ is not the final frequency of companions per primary. Some binaries may merge as they migrate within the Roche limit, and some triples may become disrupted as the outer tertiary migrates within the stability limit.  Moreover, because we remove the systems with final $M_1$~$>$~M$_{\rm 1,high}$, the selected subset with $M_1$~=[M$_{\rm 1,low}$, M$_{\rm 1,high}$] have different multiplicity statistics than the simulated population as a whole. In our baseline model, we find $f_{\rm bin}$ = 0.3 and 2.0 approximately reproduce the multiplicity statistics of solar-type and early-B primaries, respectively. In the supplementary models {\tt Mult1} and {\tt Mult2} the companion frequency is decreased or increased. 

In our baseline model, we assume fragmentation occurs with equal probability across all accretion episodes.  E.g., for our baseline solar-type model, there are 20 accretion episodes,  each of which has a $f_{\rm bin}$/20 $\approx$ 1.5\% chance of forming a new companion. The accretion rates and stochastic variability of embedded Class 0 protostars are higher than their older Class I counterparts \citep{Froebrich2006,Krumholz2009,Peters2010,Dunham2014,Hartmann2016}, and so fragmentation may preferentially occur during earlier episodes. To encompass this possibility, we consider a model {\tt Frag1} in which the probability of fragmentation is $p$ $\propto$ $t$ $^{-0.5}$, i.e., the first accretion episode at $t$ = 0.1 is three times more likely to form a new companion than the last episode. Alternatively, the accretion rate may initially increase with time or it may take several episodes for the disk to increase in mass and radius \citep{McKee2003, Girichidis2012}, and so disk fragmentation may not occur until later times (see Fig.~5 in \citealt{Kratter2008}). We therefore also consider a model {\tt Frag2} with $p$ $\propto$ $t$ $^{0.5}$ so that the likelihood of disk fragmentation is weighted toward later episodes.

Disc instabilities are stochastic and therefore can be recurrent, i.e., some discs can experience two fragmentation episodes to produce triples. In our baseline model, two fragmentation events can occur within two successive episodes. It may actually take a finite interval for the disk to recover from the initial gravitational instability, increase in mass, and become prone to fragmentation again. Nonetheless, the compact coplanar triple protostar investigated by \citet{Tobin2016} is a case example where the first companion has migrated only slightly inward to $a$~$\approx$~60~au while the outer tertiary at $a$~$\approx$~180~au (marginally stable) has only recently fragmented from the disk.

Models of gravitational disk instability suggest the initial fragment mass is $M_{\rm frag}$ $\approx$ $\epsilon \Sigma \lambda^2$, where $\Sigma$ is the disk surface density, $\lambda$ = 2$\pi H$ is the most unstable wavelength given the scale height $H$ of the disc, and $\epsilon$ $\approx$ 0.5 is a constant \citep{Goodman2004,Kratter2008,Boley2010,Kratter2016,Tobin2016}. This corresponds to $M_{\rm frag}$ $\approx$ 0.01 \msun\ and 0.1 \msun\ for representative unstable disks of solar-type and early-type primaries, respectively. For our baseline model, we adopt initial companion masses of $f_{\rm m20}$ = 0.25 fraction of mass accreted in each episode. For solar-type and B-type stars these masses are typically 0.0125 and 0.075 \msun, respectively; they are proportional to $M_{\rm tot}$. We consider half and double these companion seed masses in models {\tt CompSeed1} and {\tt CompSeed2}, respectively.
These studies also show that the initial fragment inevitably accretes and clears a gap in the disk, roughly doubling in mass. In our model, the companion growth in each accretion episode is restricted to a fraction $f_{\rm m2max}=1$ of its current mass (no more than double). This restriction is relevant only for the first episodes, when the companion has a small mass.  
We also consider supplementary models {\tt MaxAcc1} and  {\tt MaxAcc2} where the growth parameter is decreased and increased by a factor of two. 

We select the initial separation of the companions from a log-uniform distribution across the interval $a_0$ = [$a_{\rm min}$, $a_{\rm max}$].  For disks with solar-metallicity opacities, 
\citet{Moe2019} showed that the gas cooling timescales at separations below $r$~$<$~40~au are too long to achieve fragmentation. We therefore adopt a minimum initial separation for disk fragmentation of $a_{\rm min}$ = 40 au in all our models. \citet{Ansdell2016,Ansdell2018} utilized ALMA observations of CO molecular transitions to measure the gas disk radii and masses of dozens of low-mass protostars in Lupus ($\approx$1-3Myr). They found the gas radii span $r$~$\approx$~70-500 au, larger than that inferred from continuum dust emission, but with gas masses that are currently only $\approx$ 0.2 - 3 $M_{\rm J}$. Gravitational instability requires younger, more massive disks ($M_{\rm disk}$/$M_*$ $>$ 0.1;  \citealt{Kratter2016}), which are expected to be even larger. Indeed, despite their short lifetimes and propensity for instability, there are a handful of massive disks with $M_{\rm disk}$/$M_*$ $\sim$ 0.1 - 1.0 around early-B protostars with $M_*$ $\sim$ 5 - 20 \msun\ that extend to $r$~$\approx$\,500\,-\,2,000 au \citep[][references therein]{Cesaroni2007}. In our baseline model, we set the maximum separation for disk fragmentation to be $a_{\rm max}$ = 1,000 au and 3,000 au for solar-type and early-B primaries, respectively. In our model, the accretion-induced migration does not depend on the absolute value of the separation, hence the final distribution of separations and periods is a simple convolution of the initial distribution of $\log a$ with the distribution of the migration factor $\log (a_{\rm final}/a_{\rm init})$.  The choice of initial separations is almost irrelevant, if we neglect mergers.
Given the log-uniform separation distribution across $a_0$ = [$a_{\rm min}$, $a_{\rm max}$], the median disk fragmentation separation is 200 au and 350 au for solar-type and early-B systems, respectively, in our baseline model.  

\subsection{Binary Accretion}
\label{BinAcc}

Evolution of accreting binaries has  been studied by many authors over the past three  decades \citep[e.g.][]{AL1996,Bate1997}. The infalling  gas accumulates  near the  inner edge  of  the circumbinary disk,  falls  onto  the   binary  through  fast  streams (or arms),  and temporarily  settles  into  circumstellar  disks around  each  component before  being accreted by  them.  The lower-mass companion is closer to the edge of the circumbinary disk and sweeps out a larger area in its Keplerian orbit. For a binary in a circular orbit that accretes cold gas from a coplanar circumbinary disk, it is generally accepted that most of the accreted mass is directed toward the companion \citep{Bate1997,Farris2014,Young2015}. 

Let $f_2$ be the mass fraction accreted by the secondary: ${\mathrm   d}M_2  = f_2 {\mathrm  d}m $  and ${\mathrm  d}M_1 =  (1 -f_2) {\mathrm  d}m$. By definition, $f_2$ = 0.5 when the mass ratio $q$ = $M_2/M_1$ = 1 is unity.  For cold gas, i.e., negligible sound speed relative to Keplerian orbital velocity ($c$ = $c_{\rm s}$/$\sqrt{GM/a}$ $\approx$ 0.1), \citet{Farris2014}  established $f_2 = 1/(1+q)$ based on a suite of 2D hydrodynamic simulations. The secondary grows preferentially until the  components  become  comparable  in  mass.  \citet{Young2015} explored accreting binaries with different mass ratios and gas temperatures. They found a similar but slightly different relation $f_2 = q/(1+q)$ for accretion of cold gas with $c$ = 0.05. For hotter gas with $c$ = 0.25, they showed the respective accretion rates were less sensitive to the mass ratio, e.g., $f_2$ $\approx$ 0.65 for $q$ = 0.1.  For young protobinaries that have recently formed, i.e., high accretion rates and initially wide separations $a$~$\sim$~100~au ($v_{\rm orb}$ $\sim$ 3 km s$^{-1}$), the infalling gas is likely to accrete somewhere between these cold or hot regimes.  Specifically, such massive discs prone to fragmentation have mid-plane temperatures $T$~$\approx$~20\,-\,80\,K near $r$~$\sim$~100~au \citep{Boss1998,Kamp2001,Kamp2004,Kratter2006,Kratter2008,Williams2011}, yielding thermal velocities $c_{\rm s}$~$\approx$~0.3\,-\,0.6~km~s$^{-1}$ and therefore intermediate values of $c$ = $c_{\rm s}$/$v_{\rm orb}$~$\approx$0.1\,-\,0.2. \citet{Hanawa2010} and \citet{deValBorro2011} investigated the accretion of hot gas ($c$ $\approx$ 0.2) onto binaries with $q$~$>$~0.7. Spiral density patterns developed in their circumbinary discs, which caused the accretion rates onto the circumprimary and circumsecondary discs to vary with a slight preference toward the former, i.e., $f_2$ $\approx$ 0.4 - 0.5.  \citet{Zhao2013} included the effects of magnetized cores in their MHD simulations, and found the binary was fed by a collapsing pseudo-disk instead of typical circumstellar and circumbinary disks, and that the more massive primary preferentially accretes more of the gas. Finally, \citet{Dunhill2015} showed that an eccentric binary with $e$~=~0.6 precesses relative to the disk, resulting in  an oscillating $f_2$ that averages to $\approx$\,0.5 after many precession timescales. 

Infalling material onto the disk does not necessarily have the same angular momentum direction as the disk, and so each accretion episode can torque and warp the disk relative to the plane of the binary orbit. Although very close T Tauri binaries with $a$ $<$ 1 au tend to have coplanar prograde disks, the disks around wider T Tauri binaries with $a$ = 1 - 100 au exhibit a wide distribution of mutual inclinations from coplanar prograde to orthogonal \citep{Czekala2019}. Misaligned disks may tear apart into multiple rings, causing chaotic accretion onto the binary components \citep{Nixon2013,Dogan2015}. Hydrodynamic simulations show that the orbit-averaged accretion rates onto the primary and secondary are more comparable when the disk is misaligned or retrograde \citep{Hayasaki2013,Ivanov2015}. 

All the hydrodynamic simulations considered above incorporate sink particles for the binary components such that all the mass in the circumprimary and circumsecondary disks are eventually accreted by the primaries and secondaries, respectively. In reality, mass is never completely conserved. Young stellar objects lose mass and angular momenta via disk accretion winds, outflows, and jets (see reviews by \citealt{Frank2014,Bally2016,Hartmann2016}). For T Tauri stars, the measured disk mass loss rates are roughly a few percent of and linearly proportional to the mass accretion rates \citep{Hartigan1995,Rigliaco2013,Natta2014}. However, younger class 0/I protostars tend to exhibit collimated Herbig-Haro jets \citep{Reipurth2001a,Bally2016}, and are expected to have substantially lower accretion efficiencies \citep{Behrend2001,Haemmerle2016,Haemmerle2019}. To match the observed birthlines of intermediate-mass pre-MS stars on the HR diagram, \citet{Behrend2001} found that protostars accrete only $\approx$30\% of the mass from the disk. Toward high accretion rates and more massive protostars, perhaps only $\approx$10\% of the mass in the disk is accreted \citep{Haemmerle2016}. 

We predict that the accretion efficiencies will be even smaller in very young proto-binaries. Consider a 0.3\msun\ primary accreting from a marginally unstable disk, i.e. $\approx$10$^{-5}$ \msun\ yr$^{-1}$ to achieve gravitational instability.  The disk fragments and forms a new companion that accretes to $\approx$0.03\msun\ as it clears a gap and inner cavity. At this time when $q$ = 0.1, most of the mass is initially directed toward the secondary, but the secondary can only accrete and retain material on its thermal Kelvin-Helmholtz timescale $\dot{M}_{\rm KH}$ = $R L$/G$M$ $\sim$ 10$^{-7}$ \msun\ yr$^{-1}$ \citep{Kippenhahn1977,Pols1994}. In response to the high accretion rate, the secondary expands while the excess mass builds up in the circumsecondary disk. If the high accretion rate is sustained, then not even the circumsecondary disk can retain the excess material, which is either lost via polar outflows and jets, draining angular momentum from the system, or re-directed toward the circumprimary disk. In any case, when the protobinary initially forms, i.e., in the high accretion rate regime and $q$ $\lesssim$ 0.1, most of the mass is lost from the system and $f_2$ is lower than in the ideal conservative scenario. The parameter $f_{\rm m2max}$ is introduced in the model to limit the growth of low-mass companions. In our baseline model, we set $f_{\rm m2max}$ = 1, i.e., the secondary can at most double during a single accretion episode.

In our toy model, we adopt the following parameterization for the mass fraction accreted by the secondary:
\begin{equation}
f_2 = \Delta M_2/\Delta M =  0.5 + 0.5 x (1 - q)^\beta , 
\label{eq:f2mod}
\end{equation}

\noindent where $\beta$ = 0.7 in our baseline model and $x$ = [0,1] is a uniformly distributed random number uniquely generated for each accretion episode. We consider $\beta$ = 0.5 and 0.9 in models {\tt Beta1} and {\tt Beta2}, respectively. Note that $x$ = 1 represents the ideal scenario of cold gas, circular orbits, prograde coplanar discs, no magnetic fields, and mass conservation. Any departure from these five criteria cause $x$ $<$ 1, and in our toy model a uniform random variable $x = [0,1]$ for each accretion episode  accounts for these various physical processes.

\subsection{Binary Migration}
\label{BinMig}

The response of  the binary separation to the  portion of accreted gas depends on the specific angular momentum and angle of the infalling gas with respect to the binary orbit along with other factors, e.g., angular momentum losses via disk outflows and jets.  The evolution of the  binary's separation should  be proportional to the relative mass accreted by the secondary component,
\begin{equation}
\frac{ {\mathrm d} a}{a} =  - \eta \frac{  {\mathrm d} M_2} {M_2}, 
\label{eq:da}
\end{equation}

\noindent where  the  parameter  $\eta$  defines  the  speed  and  direction  of migration \citep[see e.g.  eq. 6 in ][]{Roedig2014}. In some works, the migration coefficient $\eta$ is defined in relation to the total mass increment $ {\mathrm d} m / m$. Both definitions are equivalent at large mass ratios $q$, but at small $q$ the relative mass increment of the companion, not of the total mass, is the relevant parameter for migration. 

Assuming the gas is co-moving with the binary, i.e., the gas and binary orbit have the same specific angular momentum, then accretion by the binary causes the orbit to shrink according to $a$ $\propto$ $M^{-1}$, i.e., $\eta$~=~1 \citep{Bonnell2005,Umbreit2005}. These studies also showed that if the gas is at rest with respect to the binary, then the total angular momentum $L$ of the binary orbit must be conserved, leading to $a$ $\propto$ $M^{-3}$, i.e., $\eta$~=~3. \citet{Umbreit2005} also examined counter-rotating gas with respect to the orbit, and found even stronger inward migration, i.e., $\eta$ = 5. \citet{Bonnell2005} considered a random-walk evolution, whereby the angular momentum of each infalling gas parcel is randomly oriented with respect to the binary. In this scenario, they showed analytically $a \propto  M^{-2}$, i.e., $\eta$ $=$ 2,  confirmed by their numerical simulations. Finally, \citet{Goicovic2017} used a 3D SPH code to simulate rapid accretion (within 4--8 binary periods), and measured $\eta$ = 3.45  for prograde  gas, $\eta =  5.6$ for random  orientation, and $\eta  = 7.4$  for  retrograde  gas. The random-walk or rapid-accretion scenarios likely apply to wide binaries. E.g., nascent cores at initial $a$ $\approx$ 5,000 au with $\sim$10\% their final mass will migrate to $a$ $\approx$ 50 au according to the random-walk hypothesis. Even without accretion, wide binaries can migrate significantly inward via dynamical friction \citep{Offner2010,Bate2019} (Lee et al., submitted). 

At closer separations, however, binaries likely accrete from a circumbinary disk, which has a specific angular momentum larger than the binary's. \citet{Bate1997} showed that a binary that accretes prograde and aligned gas will expand, i.e., $\eta < 0$. Utilizing a 2D grid code, \citet{Tang2017} explored whether the binary expands or shrinks as it accretes from a coplanar circumbinary disk. The answer depends on the the size of the circirumstellar disks, parameterized by  the  sink  time $\tau_s$ in units of the binary  period.  For a fast sink ($\tau_s \ll 1$),  the disks  are small, the  overall torque  is  positive, and  the binary expands.  Conversely, for slow sinks ($\tau_s = 5$), \citet{Tang2017}  showed the binary shrinks according to Eqn.~\ref{eq:da} with $\eta = 3.16$. The  gas streams arriving  at the components  are accelerated and  repelled  back  to  the  cavity edge,  and  this  ``gravitational slingshot'' mechanism brakes the  binary. \citet{Munoz2019} also performed 2D simulations of accretion from a circumbinary disk with higher resolution and over longer (viscous) timescales until a steady-state was reached. They always found orbital expansion, i.e., $\eta = -2.15$  for a circular orbit and $\eta = -0.47$ for $e=0.6$. 

There are three important caveats that would counteract the expectation from 2D hydrodynamic simulations that binaries expand as they accrete from a circumbinary disk. First, the majority of T Tauri binaries with $a$ = 1 - 100 au accrete from misaligned disks \citep{Czekala2019}. Although not a true random-walk scenario with $\eta$ = 2, the discs may be sufficiently misaligned so as to result in inward migration with $\eta$ $>$ 0. Second, the cores and disks are likely magnetized, and MHD simulations have shown that accretion of magnetically-braked material substantially shrinks the separation of the binary \citep{Zhao2013}. Finally and most important, the numerical simulations assume mass conservation, but as discussed above, most of the mass in the disks, especially the circumsecondary disk, are lost via outflows and jets, draining angular momentum from the system. 

In our toy model, we therefore consider a variable $\eta$, unique for each accretion episode, and with a positive average $\langle \eta \rangle$ $>$ 0, i.e., net inward migration. Eqn.~\ref{eq:da} applies only in the limit where d$M_2$ $<<$ $M_2$. We adopt a more general form: 

\begin{equation}
a  =  a_0 \exp ( - \eta \Delta M_2/ M_2) ,
\label{eq:a}
\end{equation}

\noindent where $\Delta M_2$ is the mass accreted by the secondary in each accretion episode. We find a uniform random variable $\eta$ = [0, 3] adequately reproduces the observations and accounts for the three factors indicated above, i.e., misaligned disks, magnetic fields, and mass loss. We consider different ranges of $\eta$ values  in supplementary models {\tt Eta1} and {\tt Eta2}, respectively. In the first accretion episodes when the companion's growth is limited by the parameter $f_{\rm m2max}$, we  evaluate Eqn.~\ref{eq:a} at the full, unlimited $\Delta M_2$ according to Eqn.~\ref{eq:f2mod} to account for the angular momentum losses and impact on the binary orbit, even though the secondary actually retains no more than $f_{\rm m2max}  M_2$ mass.

A migrating binary  may  become  too close  and  merge if the stars overfill their Roche lobes.  The  merging condition is approximately $a < 2.5 R$ \citep{Eggleton1983},  where the primary radius $R$ is  estimated  from the primary mass  according to $R/R_\odot  =   1.5  (M_1  /{\cal   M}_\odot)^{0.8}$, as appropriate for pre-MS stars. Similarly, tertiary companions may migrate within three times the separation of the inner binary, becoming dynamically unstable. In our baseline model, we eject such tertiaries (the ejected mass is lost).   In supplementary model {\tt Fold1}, we dynamically unfold such unstable configurations by increasing  the outer separation by a factor from 100 to 1000, compared to its current (unstable) separation. The unfolding factor has a log-uniform distribution. At the same, time, we shrink the inner semimajor axis by a factor of two at each unfolding.  

\subsection{Summary of the toy model}

\begin{figure}
\includegraphics[width=\columnwidth]{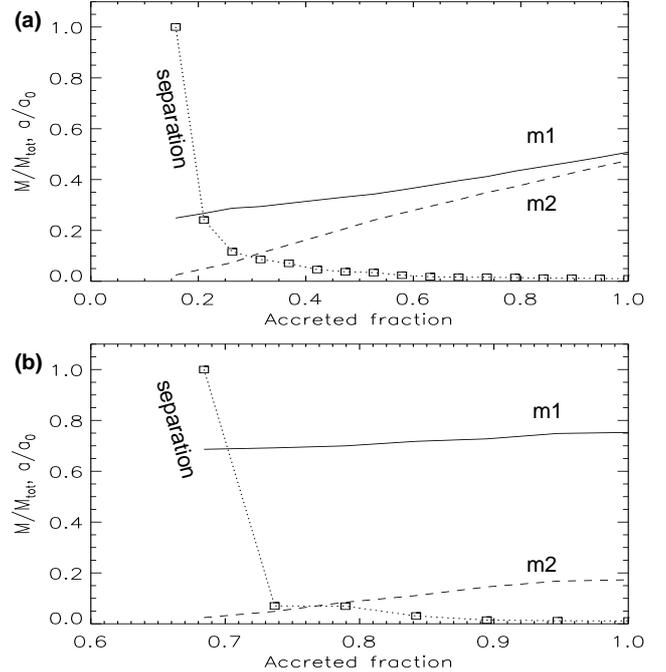}
\caption{Evolution of two simulated solar-type binary systems.  The full  and dashed lines show the fraction of the primary and secondary  mass, respectively, as a function of the total accreted mass. The squares connected by dotted line show the evolution of the semimajor  axis, also in relative units. In the first binary (a), the companion formed when 0.15 of the total mass was accreted and had a chance to become almost equal to the primary, while the separation decreased by a factor of 100. The second example (b) shows companion formation at time 0.65. In both cases, migration is strongest right after the companion formation, when its mass is still small. 
\label{fig:history} }
\end{figure}

\begin{figure}
\includegraphics[width=\columnwidth]{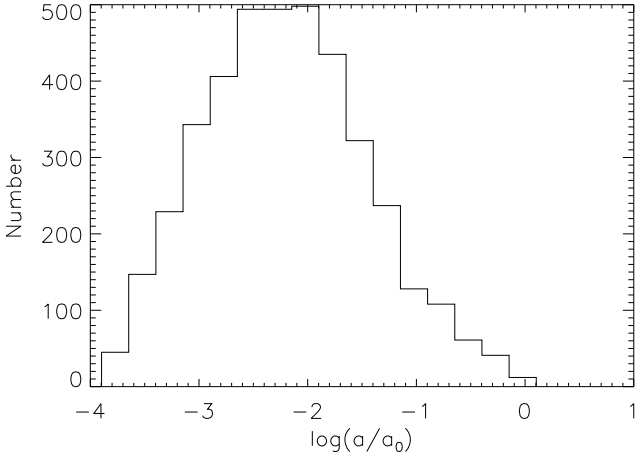} 
\caption{Distribution of the orbital migration factor  $\log (a/a_0)$ for solar-type stars.
\label{fig:migfactor} 
}
\end{figure}

The toy model generates large samples of simulated binaries by implementing the physically-motivated  prescriptions discussed above in a simple numerical code. Here its operation is summarized. The baseline parameters of the model for solar-type and B-type stars are listed in Table~\ref{tab:par}. 

For  each  simulated  system,  the total  mass $M_{\rm tot}$, uniformly  distributed between $M_{\rm tot,0}$ and $3 M_{\rm tot,1}$, is generated. This defines the initial mass of the primary star,  $f_{m0} M_{\rm tot}$, and the mass accreted in each episode, $m_{\rm acc} = M_{\rm tot}/K_{\rm step}$. The mass is added to the primary star until a companion with the initial (seed) mass of $f_{\rm m2} m_{\rm acc}$ is formed. The probability of companion formation in each episode is $f_{\rm bin}/K_{\rm step} $, and its initial separation is chosen from a log-uniform distribution in the interval [$a_0$, $a_1$]. Formation of additional companions in subsequent accretion episodes is not prohibited, but only dynamically stable outcomes are allowed (companion separation more than three times larger than the separation of the inner binary).  

The evolution of a system (single, binary, triple, etc.) in response to the an accretion episode is implemented as a subroutine.  The accreted mass  $m_{\rm acc}$ is distributed between components according to the Eqn.~\ref{eq:f2mod} using the parameter  $\beta$ and a uniformly   distributed random number $x$, generated independently for each episode. The growth of the companion is restricted by the parameter $f_{\rm m2max}$. Then Eqn.~\ref{eq:a} is applied, with a random $\eta$ uniformly distributed between $\eta_1$ and $\eta_2$, independently  in each episode. Using the updated separation, we check the merging condition (if satisfied, the binary becomes again a single star with the sum of component's masses). If the number of  components exceeds two, the evolution subroutine is called recursively (with independent random numbers), and the condition for dynamical stability is checked to eliminate (or unfold) unstable triples.  Figure~\ref{fig:history} illustrates evolution of two typical solar-mass binaries. 

When all mass is accreted by the system in $K_{\rm step}$ episodes, we  consider only single stars, binaries, and higher-order hierarchies with primary masses in the  $M_{\rm tot,0}$, $M_{\rm tot,1}$ range. Single stars with these masses are simply counted. The counters of mergers and disruptions are incremented only for systems in the requested range of primary mass.  Generation of systems is repeated until the total requested number of binaries is reached. Figure~\ref{fig:migfactor} shows the distribution of the ratio of final and initial semimajor axes $\log (a/a_0)$ (i.e. the migration factor) for solar type stars. The median is $-2.2$, meaning that the initial binary separation is reduced typically by two orders of magnitude. In the toy model, the distribution of final binary separations or periods is a simple convolution of the initial distribution (assumed log-uniform) with that of the migration factor.  

\begin{table}
\center
\caption{Statistical parameters}
\label{tab:stat}
\begin{tabular}{l l} 
\hline
Parameter   & Description   \\
\hline
CF & Companion fraction, $N_{\rm comp}/N_{\rm sys}$ \\
BF & Fraction of pairs, $\log P < \log P_{\rm max}$ \\
TF &  Fraction of triples, $\log P_{\rm out} < \log P_{\rm max}$ \\
$f_{\rm disrupt}$ & Fraction of disrupted systems  \\
$f_{\rm merge}$ & Fraction of mergers  \\
$\gamma_q$ & Power-law of $f(q)$, $0.3 < q < 0.95$ \\
$f_{\rm twin}$ & Excess twin fraction among  $q>0.3$ \\ 
$\gamma_P$ & Slope of $N(\log P)$ in the (0.3, 2) interval \\
\hline
\end{tabular}
\end{table}

The statistics of the simulated binary population is characterized by several parameters  (Table~\ref{tab:stat}). We consider all pairs, independently of their hierarchy (both inner and outer subsystems in triples) and determine the total companion fraction CF accordingly. However, the binary and triple fractions, BF and TF, refer only to pairs with periods less than  $P_{\rm max}$, chosen at $10^4$ days for solar-type stars and $10^{3.4}$ days for B-type stars. Only short-period binaries can be generated by the toy model, so the resulting statistics are meaningful at separations below  $\sim$10 au, motivating the choice of  $P_{\rm max}$.  The mass ratio distribution at $q> 0.3$ is characterized by the truncated power law with index $\gamma_q$ and the excess twin fraction $f_{\rm twin}$ (fraction of binaries with $q>0.95$ in excess of the power law, relative to all binaries with $q>0.3$), as in \citet{Moe2017}. We compute these parameters in two period intervals, $\log (P)$  (in days) of (0 - 2) and  (2 - 4) for solar-type binaries, (0 - 1.7) and (1.7 - 3.4) for B-type binaries (with  subscripts S and L, respectively).   The period distribution $ N(\log P)$ is fitted by a straight line in the $\log(P)$  interval  (0.3 - 2), and its slope $\gamma_P$ quantifies the overall migration. Strong migration of B-type binaries produces an excess of short periods, with a negative $\gamma_P$ and a large rate of mergers. In contrast, for solar-type binaries the slope is positive and the number of mergers is very small. 

\section{Results}
\label{sec:res}

We list the results of our simulations for solar-type and early-B binaries in Tables~\ref{tab:sunpar} and \ref{tab:bpar}, respectively, for our baseline and all supplementary models. We also compare the simulated parameters to the observations according to the 67-pc sample of solar-type systems \citep{Tok2014}, meta-analysis of solar-type and early-type multiples \citep{Moe2017}, and other surveys as described below. Data   of  recent publications on these stars are used. The observed statistical parameters of binaries are given in the first lines of both Tables in italics. 

\subsection{Solar-mass binaries}
\label{sec:sun}

\begin{figure}
\includegraphics[width=\columnwidth]{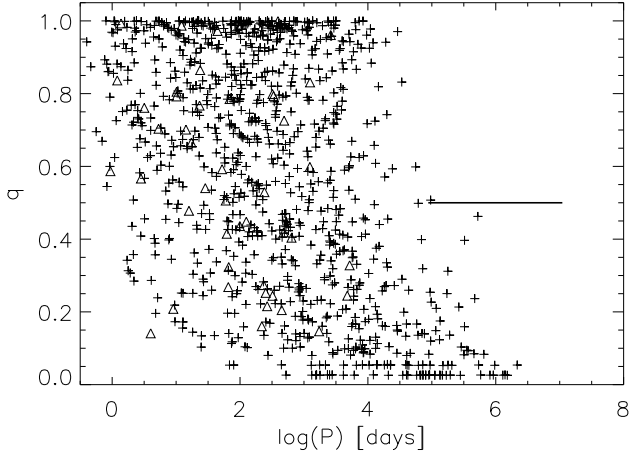} 
\caption{$P,q$  plot  of 1000  simulated  solar-mass  binaries.  The
  horizontal  line  shows  the  range  of  initial  periods.  Binary,
  triple, and quadruple systems are plotted as pluses, triangles, and
  squares. 
\label{fig:sun_pq} 
}
\end{figure}

In Fig.~\ref{fig:sun_pq}, we show the periods $P$ and mass ratios $q$ of individual companions to solar-type primaries in our baseline model. The thick line depicts the range of initial binary periods, corresponding to $a_0$ = 40\,-\,1,000 au. Binaries that formed early have, on average, migrated to shorter periods and have larger $q$. Various surveys have demonstrated that solar-type primaries exhibit a dearth of close brown dwarf companions within $P$ $\lesssim$~100 days, commonly known as the brown dwarf desert, but that the frequency of brown dwarf companions increases with increasing separation \citep{Grether2006,Kraus2008,Kraus2011,Csizmadia2015,Wagner2019,Shahaf2019}. Our model naturally reproduces the brown dwarf desert, as it is very difficult for very low-mass companions to migrate within $P$~$\lesssim$~100~days without also accreting above $M_2$ $>$ 0.08\,\msun. Meanwhile, we simulate many brown dwarfs at longer periods, which were companions that fragmented at late times and therefore accreted and migrated very little.

\begin{figure} 
\includegraphics[width=\columnwidth]{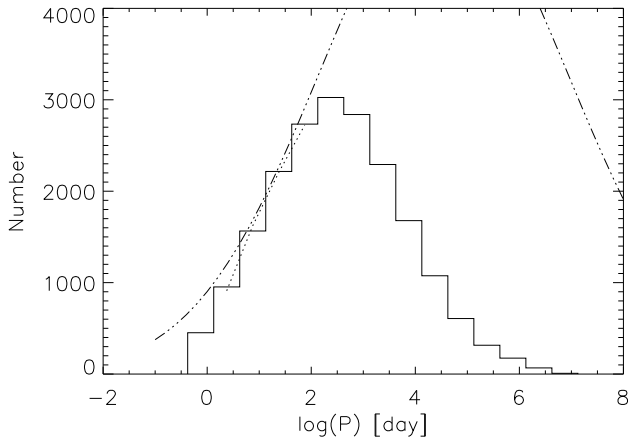} 
\caption{Period distribution   of  solar-type binaries. Full line --- simulations, dash-dot line --- Gaussian model, dotted line --- power law.
\label{fig:sun_p} 
}
\end{figure}

In the field, $\approx$\,20\% of solar-type stars have MS companions within $a$~$<$~10~au ($P$ $<$ 10$^4$ days; \citealt{Moe2017,Moe2019}). This fraction increases slightly to BF~$\approx$~22\% after including the observed population of brown dwarf companions across $a$ = 1\,-\,10 au. In the simulated sample, we get a slightly larger BF of 30\%, but it is easily adjustable by reducing  $f_{\rm bin}$. 

The overall triple-star fraction of solar-type stars is $\approx$\,13\%, but only 17 ($\approx$0.3\%)  of the solar-type primaries in the 67-pc sample are in known compact triples with $a_{\rm out}$ $<$ 10 au \citep{Tok2014}. However, the observed compact triple star fraction is a lower limit due to incompleteness, as it is very difficult to detect faint low-mass tertiaries within $a_{\rm out}$~$<$~10~au. Indeed, in our baseline model, the majority of compact triples have $q_3$ = $M_3$/$M_1$ $<$ 0.3 (see Section~\ref{sec:triple}). \citet{Brandt2018} recently combined {\it Gaia} DR2 and {\it Hipparcos} astrometry to detect accelerating systems, indicative of binary companions with intermediate separations of $a$~$\approx$~1\,-\,50 au. Of the 232  binaries with $P$~$<$~100~days in the 67-pc sample, 31 exhibit statistically significant astrometric acceleration. These accelerations are, mostly, not caused by wide, easily resolvable tertiaries beyond $a_{\rm out}$~$>$~10~au. Instead, the majority of the 31 close binaries exhibiting astrometric acceleration are likely previously unrecognized compact triples with $a_{\rm out}$~$<$~10~au. Moreover, it is also challenging to identify compact A-Ba,Bb triples, e.g., a pair of low-mass M-dwarfs closely orbiting a solar-type primary, utilizing traditional methods of RV monitoring and high-contrast imaging. Nonetheless, examinations of {\it Kepler} eclipsing binaries exhibiting eclipse timing variations reveal a large population of compact A-Ba,Bb triples \citep[][references therein]{Borkovits2016}. Considering the selection biases, we estimate a compact triple star fraction of TF $\approx$ 1\% for solar-type primaries, similar to our baseline model value of TF~$=$~3\%. 

Although we selected $f_{\rm bin}$ = 0.3 apriori to roughly match the observed close companion fraction, the {\it ratio} of compact triples to close binaries was not predetermined. We estimate TF/BF $\approx$ 1\%/22\% $\approx$ 5\% for the observed sample of solar-type primaries, and compute TF/BF = 3\%/30\% $\approx$ 10\% for our baseline model. By treating disc fragmentation as a stochastic repeatable process, our model qualitatively reproduces the observed ratio of compact triples to close binaries.

Both the observed and simulated solar-type binaries are weighted toward larger periods, as shown in Figure~\ref{fig:sun_p}. In our baseline model, only 0.4\% of solar-type primaries had merged with a very close binary companion ($\log P$ $\lesssim$ $-$0.4), and  4\% of primaries have companions that migrated to $P$~=~1\,-\,10 days. The latter value is  comparable to 2\% observed both in the field and in young star-forming regions \citep{Moe2018}. The field solar-type binary period distribution is roughly log-normal, peaking at $\log P$ = 4.8 with dispersion of 2.3 dex \citep{Duquennoy1991,Raghavan2010,Tok2014}. Across $\log P$ = 0.3\,-\,2, the observed slope is $\gamma_{\rm P}$ = d$N$/dlog$P$ = 0.7, consistent with the simulated value of $\gamma_{\rm P}$. However, at longer periods $\log P$ $\gtrsim$ 2.5, the simulated distribution flattens and turns over, underestimating the true frequency of companions at intermediate separations (Fig.~\ref{fig:sun_p}). 

We surmise that binaries which formed via core fragmentation, which are not included in our model, begin to contribute at a non-negligible level beyond $a$~$\gtrsim$~1~au. Indeed, although the majority of double-double quadruples have been observed in loose hierarchies with $a_{\rm out}$ $>$ 10 au, a few have been detected in compact configurations with $a_{\rm out}$ = 1\,-\,10 au \citep{Tok2014}. Such double-double quadruples cannot derive from successive inside-out disk fragmentation episodes as encapsulated by our model, but instead via an outside-in process whereby a wide binary first forms via core fragmentation and then both of those components subsequently split via disk fragmentation. The existence of compact double-double quadruples with $a_{\rm out}$ = 1\,-\,10 au strongly suggests that binaries that formed via core fragmentation can migrate to such intermediate separations. Core fragmentation binaries become more relevant with increasing separation. The contribution from disk versus core fragmentation binaries are comparable near $a$ $\approx$ 50 au (see Introduction), the peak in the overall period distribution of solar-type binaries. Both \citet{Moe2019} and \citet{ElBadry2019} showed that the fraction of wide binaries ($a$ $>$ 200 au; $P$~$>$~10$^6$~days) is independent of metallicity, and so they concluded nearly all such wide binaries are the result of core fragmentation. As shown in Fig.~\ref{fig:sun_p}, our model population of disk fragmentation binaries steadily declines to zero near log\,$P$ = 6, consistent with the observational constraints.

\begin{figure} 
\includegraphics[width=\columnwidth]{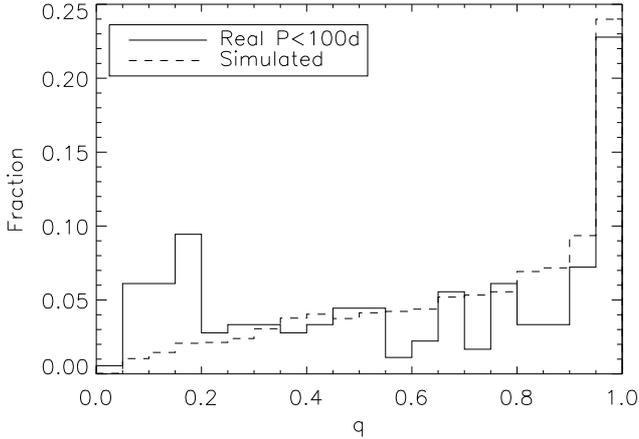} 
\caption{Distribution  of the  mass ratio  $q$ of  solar-type binaries  with $1 <  P <100$ days. Full line --- binaries  in the 67-pc sample   \citep{Tok2014}, dashed  line --- simulations.   
\label{fig:sun_q} 
}
\end{figure}

Our model also reproduces the observed excess fraction of twins with $q$ $>$ 0.95, especially evident at short periods. In Figure~\ref{fig:sun_q}, we compare the simulated mass-ratio distribution of solar-type binaries with $P$ = 1\,-\,100 days to the observed distribution in the 67-pc sample \citep{Tok2014} across the same period range. For the latter, we use the measured mass ratios of double-lined spectroscopic binaries (SB2s) and the minimum mass ratios of SB1s. Across the interval $q$ = 0.3\,-\,1.0, the observed and simulated mass-ratio distributions are well described by a power-law with slope $\gamma_{q,S} \approx 1$ and an excess twin fraction $f_{\rm twin,S} \approx 0.2$   relative to the power-law component. However, \citet{Moe2017} noted that a non-negligible fraction of solar-type SB1s with small $q$ in the field contain white dwarf companions, and so the intrinsic distribution of close MS companions are further weighted toward higher $q$. Accounting for this bias, we estimate $\gamma_{q,S}$ = 0.8 and $f_{\rm twin,S}$ = 0.24 in the field. Fitting the simulated sample of solar-type binaries with $P$ = 1\,-\,100 days yields $\gamma_{q,S}$ = 1.5 and $f_{\rm twin,S}$ = 0.20, qualitatively consistent with the observations. 

With increasing period, the mass-ratio distribution becomes weighted toward smaller values. In our simulations, the power-law slope decreases from $\gamma_{q,S}$ = 1.5 to $\gamma_{q,L}$ = 0.2 from $\log P$~(days) = 0\,-\,2 to 2\,-\,4, and the excess twin fraction also decreases from $f_{\rm twin,S}$ = 0.20 to $f_{\rm twin,L}$ = 0.15. There is strong observational evidence that the solar-type excess twin fraction decreases across $\log P$ = 0\,-\,4 \citep{Lucy1979,Tok2000,Moe2017}. Utilizing the 25-pc \citep{Raghavan2010} sample of solar-type primaries, \citet{Moe2017} measured the excess twin fraction to decrease linearly with $\log P$ such that $f_{\rm twin,L}$/$f_{\rm twin,S}$~$\approx$~0.7, consistent with the simulated ratio $f_{\rm twin,L}$/$f_{\rm twin,S}$~$\approx$~0.75.

\subsection{Massive binaries}
\label{sec:B}

\begin{figure} 
\includegraphics[width=\columnwidth]{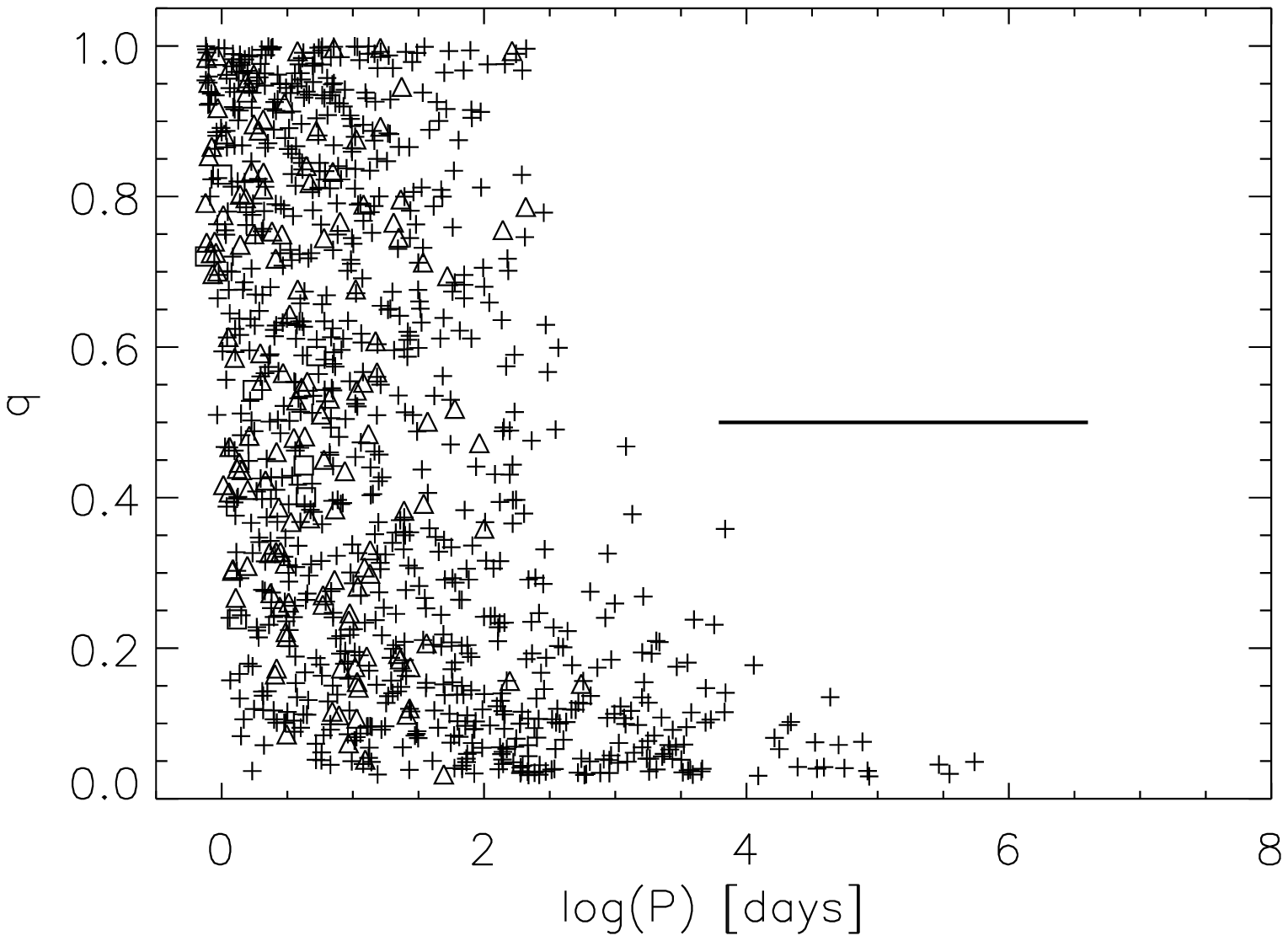} 
\caption{$P,q$  plot  of 1000  simulated  B-type  binaries.  The  horizontal  line  shows  the  range  of  initial  periods.  Binary,  triple, and quadruple systems are plotted as pluses, triangles, and  squares. 
\label{fig:B_pq} 
}
\end{figure}

\begin{figure} 
\includegraphics[width=\columnwidth]{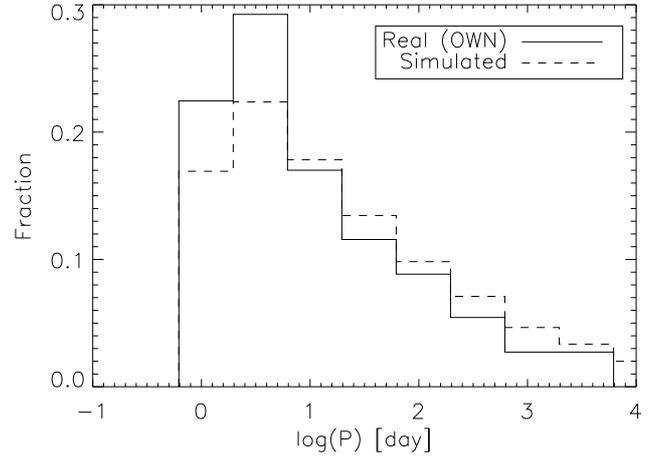} 
\caption{Histograms of periods  for  B-type binaries. The observed period distribution of massive binaries  is taken from the OWN survey by \citet{Barba2017}.  
\label{fig:B_hist} 
}
\end{figure}

\begin{figure} 
\includegraphics[width=\columnwidth]{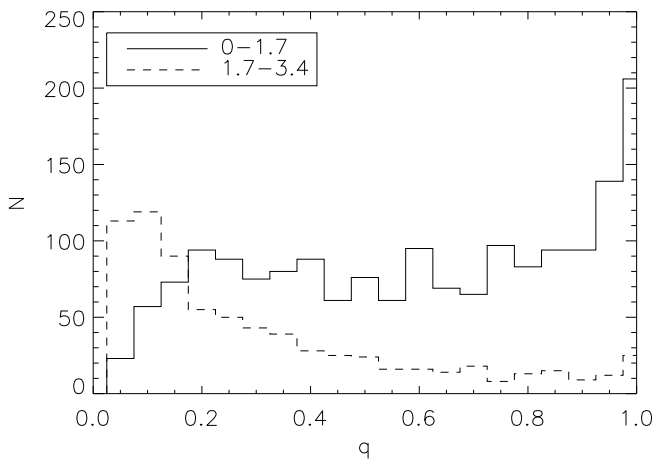} 
\caption{Histogram of the mass ratio of simulated B-type binaries in period intervals $\log(P)$ of  0-1.7 and 1.7-3.4.
\label{fig:B_q} 
}
\end{figure}

Massive stars form by accreting  more matter than solar-type stars. We set a larger number of accretion bursts $K_{\rm step}=50$ accordingly, as   well  as   a   larger  probability   of   forming  a   companion. Figures~\ref{fig:B_pq}   shows  the  results  for B-type binaries. For volume-limited samples, $\approx$50\% of early-B stars have companions within $a$~$<$~10~au ($\log P$~$\lesssim$~3.2), and $\approx$10\% of early-B primaries are in compact triples with $a_{\rm out}$ $<$ 10 au \citep{Moe2017}. In our baseline early-B model with $f_{\rm bin}$ = 2.0, we simulate BF = 0.56 and TF = 0.13, respectively,  close to the observations. The simulated ratio of compact triples to close binaries, TF/BF $\approx$ 0.2, is higher for early-B systems, consistent with the observations. In our model, disk fragmentation is a Poisson process, and so increasing the mean number of companions $f_{\rm bin}$ not only increases the ratio of close binaries to single stars but also the ratio of compact triples to close binaries and the frequency of disrupted unstable triples.

Both the observed and simulated period distributions of early-type binaries are skewed toward short periods, as shown in Fig. ~\ref{fig:B_hist}. The formation of such OB binaries requires massive disks and significant accretion, which implies very efficient migration. In our baseline model, 33\% of the early-B primaries merged with a companion, creating a cliff in the period distribution near $\log P$~$\approx$~$-$0.2. Toward longer periods, the simulated early-B binaries follow a slope $\gamma_{ P}$ = $-$0.4, i.e. a decrease in frequency with $\log P$. The observed samples of close spectroscopic binary companions to O-type stars \citep{Sana2012,Barba2017} and mid-B stars \citep{Abt1990} yield slopes $\gamma_{\rm P}$ $\approx$ $-$0.5 and 0.0 (\"Opik's law), respectively. The observed period distribution of early-B eclipsing binaries provide $\gamma_{\rm P}$ $\approx$ $-$0.2 after correcting for selection effects \citep{Moe2013}, halfway between the spectroscopic O-type and mid-B samples and matching the simulations. 

Analysis of binary statistics from \citet{Moe2017} in the full range of primary mass $M_1$ leads to the approximate dependence of the slope on mass as $\gamma_P \approx 0.7 - 0.9 \log M_1$. When we change only the primary mass range in  B-type star simulations, the resulting parameter  $\gamma_{P}$ decreases with increasing mass, e.g.  $\gamma_{P}=-0.15$ for [2-4] \msun ~and  $\gamma_{P}=-0.50$ for [20-40] \msun. However, the period distribution is very sensitive to other parameters such as $\eta$, which can be tuned to match the observations.

As shown in Fig.~\ref{fig:B_pq}, the simulated early-B binaries also exhibit an anti-correlation between $P$ and $q$, but in a manner that is different from solar-type binaries and consistent with the observed population of massive binaries. For example, there is no deficit of close, extreme mass-ratio companions to early-B primaries in our simulations, unlike the brown dwarf desert observed for solar-type primaries. \citet{Moe2015a} discovered several eclipsing low-mass pre-MS companions to early-B MS primaries with very short periods $P$~$<$~10~days. They estimated the occurrence rate of very close $q$ = 0.05\,-\,0.15 companions to massive stars is similar to those with $q$~=~0.15\,-\,0.25, consistent with our simulations.  

For early-type binaries, the power-law slope $\gamma_{q}$ decreases substantially across $\log P$ = 0\,-\,3.2, much more so than for solar-type binaries (Fig.~\ref{fig:B_q}). In our baseline early-B model, we compute $\gamma_{q,S}$ = 0.3 and $\gamma_{q,L}$ = $-$1.3 for short and long periods, respectively. Although very close companions to OB primaries approximately follow a uniform mass-ratio distribution \citep{Sana2012,Kobulnicky2014}, there is a large body of evidence that early-type binaries with intermediate separations are substantially skewed toward small mass ratios $q$~$\approx$~0.3 \citep{Rizzuto2013,Moe2015b,Gullikson2016,Moe2017,Murphy2018}. Based on these various surveys, \citet{Moe2017} estimated $\gamma_{q,L}$ = $-$1.5 for early-B binaries, similar to the results of our baseline model.

The excess twin fraction is substantially reduced for close early-B binaries and quickly diminishes with increasing separation. In our baseline model, we measure $f_{\rm twin,S}$ = 0.09 and $f_{\rm twin,L}$ = 0.06 across short and long periods, respectively. Based on a compilation of early-type spectroscopic \citep{Sana2012,Kobulnicky2014} and eclipsing \citep{Pinsonneault2006,Moe2013} binaries, \citet{Moe2017} estimated $f_{\rm twin,S}$~$\approx$~0.08\,-\,0.15, depending on primary mass, across $P$ = 2\,-\,20 days. These values are consistent with the simulations and considerably smaller than that measured for close solar-type binaries. Meanwhile, toward longer periods $P$~$\gtrsim$~20~days, \citet{Moe2017} measured the excess twin fraction of OB binaries to be consistent with zero, placing an upper limit of $f_{\rm twin,L}$~$\lesssim$~0.05, similar to the value in our baseline model.

\subsection{Triple systems}
\label{sec:triple}

\begin{figure} 
\includegraphics[width=\columnwidth]{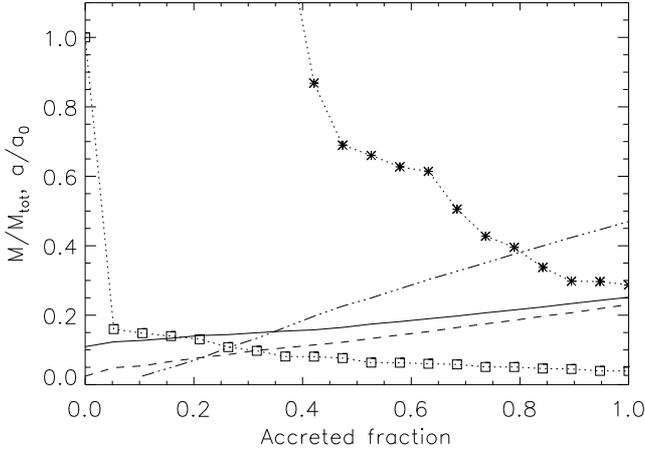} 
\caption{Example of a triple star evolution. As in Figure~\ref{fig:history}, the masses of the primary and secondary stars in the inner binary, formed in the first accretion episode, are plotted by the full and dashed lines. The tertiary component (dash-dot line) formed at time  0.1 with a semimajor axis of 200 au, much wider than the initial inner binary (outside the plot limit).  The final masses of the stars are  0.54,   0.50, and 1.0 \msun. The inner and outer separations, relative to the initial inner separation, are plotted by squares and asterisks.
\label{fig:trip0} 
}
\end{figure}

Hierarchical systems can be formed  in several different ways, e.g. by core fragmentation for the  outer subsystem and disk fragmentation for the  inner  subsystem(s). Our  model  considers  only  one process,  disk fragmentation,  and therefore  is not  expected to  reproduce  the full range  of  real hierarchies.   Its  relevance  is  limited to  compact hierarchies with  outer separations less  than $\sim$50 au  where even the  outer  subsystem can  be  a  product  of disk  fragmentation  and migration.  Such triple systems  form ``from inside out'', by adding outer companion to the existing pair and subsequent migration. As stated above, this scenario cannot explain the 2+2 quadruple systems, which must originate in a different way. 

Although triple systems are more frequent among massive stars, their statistics and distribution of hierarchical configurations are well established observationally only for solar-type stars, discussed in this Section. Figure~\ref{fig:trip0} illustrates the evolution of a simulated  solar-type triple system where the tertiary component has formed early and accreted most of the mass, forming a double twin (both inner and outer mass ratios are close to one). However, this case is atypical because most triple systems form late and their outer companions are less massive compared to both inner stars.

\begin{figure} 
\includegraphics[width=\columnwidth]{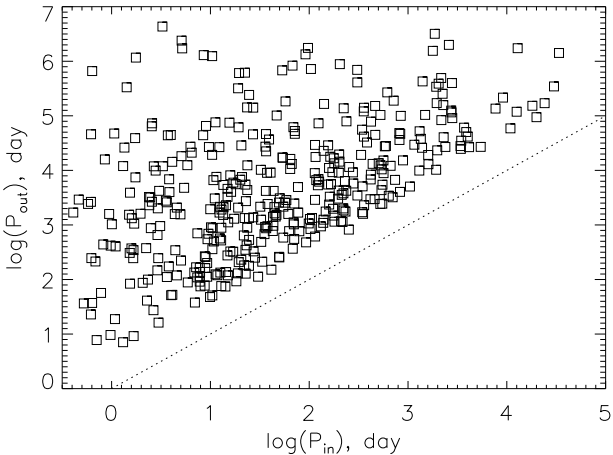} 
\caption{Triple systems with solar-type primaries: periods of  inner and outer subsystems (the dotted line marks equality).
\label{fig:sun_trip} 
}
\end{figure}

Figure~\ref{fig:sun_trip}  shows the hierarchical distributions of simulated  triple systems with solar-type primaries.  The lower envelope of points delineates the adopted  crude dynamical stability limit  (separation ratio $>$3). Note  the relatively  empty  lower-left corner,  i.e.  the paucity  of triple systems with very short outer periods $P_{\rm out}$ $<$ 100 days. These extremely compact triples  must have formed very early, and the difficulty in surviving the migration evolution without becoming dynamical unstable leads to their relative rarity.

\citet{Tok2006} demonstrated that the closest binaries are most likely to be in hierarchical triples. After correcting for incompleteness, they reported $\approx$\,96\% and $\approx$\,68\% of binaries with $P$ = 1\,-\,3 days and $P$ = 3\,-\,6 days, respectively, have outer tertiaries. The majority of the observed tertiaries to very close binaries are not compact. For example, only 5/53 = 9\%\,$\pm$\,4\% of binaries with $P$~=~1\,-\,5~days in the 67-pc sample have compact tertiaries within $P_{\rm out}$~$<$~10$^4$~days, consistent with the value of 13\% computed in our baseline model. As discussed in the Introduction, some studies have interpreted the \citet{Tok2006} observations as evidence for hardening the closest binaries via Kozai-Lidov cycles in misaligned triples coupled to tidal friction \citep{KCTF,Fabrycky2007,Naoz2014}, while others have concluded that the majority of very close binaries derive solely from disk migration \citep{Moe2018}. Indeed, our toy model naturally produces very close binaries via disk migration alone. Given the same primary mass, triple star formation requires more mass and accretion, on average, to achieve two fragmentation episodes. The additional mass and accretion facilitates in the inward migration of the inner binary, explaining the anti-correlation between binary period and triple star fraction.

\begin{figure} 
\includegraphics[width=\columnwidth]{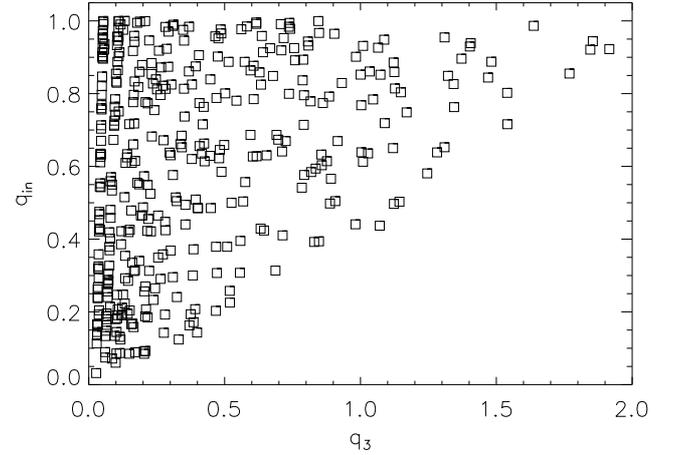} 
\caption{Mass ratios  in the outer  and inner subsystems  of simulated solar-type triple systems.
\label{fig:sun_qq} 
}
\end{figure}

Figure~\ref{fig:sun_qq} shows the  mass ratios of simulated solar-type triples. The outer  mass ratio $q_3 = M_3/M_1$  is defined relative to the inner  primary component,  rather than to  the total  inner binary mass,     as    this    parameter     is    easier     to    determine from observations.\footnote{The traditional mass ratio $q_{\rm out}= M_3/(M_1 + M_2)$   is related to $q_3$ as $q_3 = M_3/(M_1 + M_2)(1 + M_2/M_1) = q_{\rm out}(1 + q_{\rm in})$. }  The plot can be compared to  Figure~5 in \citet{Tokovinin2008}.  In systems with $q_3<1$, the most massive star belongs to the inner binary (hierarchy of Aa,Ab--B type), and in systems with $q_3>1$ the tertiary is more massive (hierarchy of A--Ba,Bb type). Only $\approx$\,12\% of simulated triples have  $q_3>1$. Rare double twins, like the one in Fig.~\ref{fig:trip0}, have $q_3 \approx 2$ and are located in the upper right corner of the plot in Figure~\ref{fig:sun_qq}. The median value of $q_3$  is 0.19,  smaller that  0.39 in the  real triple  stars. The latter, however, is positively biased by observational selection which disfavors discovery of low-mass tertiary companions (see above). Simulated triples with  massive tertiaries, like the one in Fig.~\ref{fig:trip0}, formed early and migrated more.  Consequently, we note an anti-correlation between $P_{\rm out}$ and $q_3$ for the outer tertiaries, similar to the anti-correlation between $P$ and $q$ for the inner binaries. Overall, the tertiaries are weighted toward small mass ratios $q_3$~$<$~0.3 and do not exhibit an excess of twins. According to our model, if a twin binary within $a$~$<$~10~au is observed, than it is unlikely that either of the components will have their own subcompanions. 

The lower-right corner  of the plot is empty, meaning  that there  are no  simulated triples  where both secondary and tertiary components have small masses. In our model, triple stars which formed inside-out by cascade of disk fragmentations with subsequent migration must have $q_{\rm in} \gtrsim 0.5  q_3$ or, equivalently, $M_3 \lesssim 2 M_2$. However, in the \citet{Borkovits2016} sample of compact triples, where the ratio of eclipse depths indicates the mass ratio $q_{\rm in}$ of the inner eclipsing binary while eclipse timing variations provide the masses $M_1$+$M_2$ and $M_3$, there are a handful of A--Ba,Bb triples with $q_3$~$>$~2. For example, KOI-126 (KIC 5897826) contains a triple eclipsing nearly coplanar A--Ba,Bb triple with $q_3$ = 5.6 in which a very close pair of low-mass M-dwarfs with $M_{\rm Ba}$ = 0.24\,\msun, $M_{\rm Bb}$ = 0.21\,\msun, and $P_{\rm in}$ = 1.78 days orbits a G1\,IV primary with $M_{\rm A}$ = 1.35 \msun\ at $P_{\rm out}$ = 33.9 days \citep{Carter2011}. The inner binary likely formed at very early times, and therefore migrated toward $P_{\rm in}$ $\approx$ 2 days while evolving into a near twin with $q_{\rm in}$ $\approx$ 0.8, and then the outer tertiary fragmented, accreted most of the remaining mass to $M_{\rm A}$ = 1.35\,\msun, and migrated to $P_{\rm out}$ $\approx$ 34 days. Our model cannot produce such A--Ba,Bb triples with $q_3$ $>$ 2 because we assume most of the mass is accreted by the inner binary if its combined mass is less than that of the tertiary. In any case, only a small minority of compact triples in the \citet{Borkovits2016} sample have $q_3$~$>$~2, most of which have $P_{\rm in}$ $<$ 5 days and $M_{\rm A}$ $>$ 1.3\msun\, (not true solar-type primaries per our definition). The majority of their solar-type triples have large $q_{\rm in}$ and small $q_{\rm 3}$, consistent with the parameter space of our models as shown in Fig.~\ref{fig:sun_qq}.

\subsection{Variation of the parameters}
\label{sec:par}

We  have chosen the default parameters of  the  toy  model (Table~\ref{tab:par})  to mimic the real binary  statistics, at least qualitatively (one cannot expect a perfect match from our crude model). Here we explore how changes of the parameters affect the outcome of the simulations. Supplementary models with these changes were introduced above in Section~\ref{sec:model}. The statistical parameters are described in Table~\ref{tab:stat}. Table~\ref{tab:sunpar} shows the statistics of simulated solar-type binaries resulting from supplementary models. The default simulations were performed for 20,000 binaries. Then we change one parameter at a time, generate 4,000 binaries, and determine their statistics. The last column shows the modified parameter values and, in brackets, its default value. Table~\ref{tab:bpar} gives similar information for B-type stars. For reference, the observed parameters are given in the first line of both tables.

\begin{table*}
\center
\caption{Supplementary models for solar-type stars}
\label{tab:sunpar}
\begin{tabular}{l ccc cc  cc cc c l} 
\hline
Model &  CF & BF & TF & $f_{\rm disrupt}$ & $f_{\rm merge}$ &  $\gamma_{q,S}$ & $f_{\rm twin,S}$  & $\gamma_{q,L}$ & $f_{\rm twin,L}$  & $\gamma_P$ & Comment \\
\hline
Observed & \ldots & {\it 0.24} & {\it 0.02} & \ldots & \ldots &  {\it 0.8}  & {\it 0.24}  &  {\it 0.3}  & {\it 0.16}  &  {\it 0.7} &   \\
\hline
Baseline & {\bf 0.37} &{\bf 0.30} & {\bf 0.03} & {\bf 0.03} & {\bf  0.00} &   {\bf 1.5}  & {\bf 0.20}  &  {\bf 0.2}  & {\bf 0.15}  &  {\bf 0.71} & Default  \\
PrimSeed1& 0.39 &0.32 &0.03 &0.03 &0.01 &  1.5 & 0.29 &    0.2 & 0.25 &  0.55 & $f_{\rm m0}=0.05$ (0.1)  \\  
PrimSeed2& 0.38 &0.30 &0.03 &0.03 &0.00 &  1.4 & 0.06 &    0.2 & 0.04 &  0.74 & $f_{\rm m0}=0.15$ (0.1) \\ 
CompSeed1& 0.37 &0.22 &0.02 &0.03 &0.00 &  6.0 & 0.63 &    1.4 & 0.21 &  4.11 & $f_{\rm m20}=0.5$ (0.25) \\ 
CompSeed2& 0.16 &0.14 &0.01 &0.02 &0.16 &  0.5 & 0.08 &   $-$0.5 & 0.12 & $-$0.18 & $f_{\rm m20}=0.1$ (0.25) \\ 
MaxAcc1  & 0.39 &0.27 &0.02 &0.03 &0.00 &  2.4 & 0.33 &    1.2 & 0.18 &  2.13 & $f_{\rm m2max}=2.0$ (1.0) \\ 
MaxAcc2  & 0.28 &0.24 &0.02 &0.03 &0.08 &  0.4 & 0.11 &    0.0  & 0.06 & $-$0.09 & $f_{\rm m2max}=0.5$ (1.0) \\ 
Eta1     & 0.38 &0.21 &0.02 &0.02 &0.00 &  1.7 & 0.14 &    0.9 & 0.15 &  1.33 & $\eta=[-1,3]$ ([0,3]) \\     
Eta2     & 0.35 &0.31 &0.03 &0.04 &0.02 &  1.2 & 0.14 &   $-$1.4 & 0.10 &  0.00 & $\eta=[1,3]$  ([0,3])   \\ 
Step1    & 0.39 &0.28 &0.02 &0.03 &0.00 &  1.9 & 0.24 &    0.5 & 0.20 &  1.22 & $K_{\rm step}=10$ (20)\\ 
Step2    & 0.37 &0.31 &0.02 &0.04 &0.02 &  1.3 & 0.13 &    0.0 & 0.08 &  0.30 & $K_{\rm step}=40$ (20)\\ 
Mult1    & 0.27 &0.22 &0.01 &0.02 &0.00 &  1.4 & 0.18 &    0.4 & 0.12 &  0.74 & $f_{\rm bin}=0.2$ (0.3) \\ 
Mult2    & 0.57 &0.43 &0.06 &0.08 &0.01 &  1.3 & 0.18 &    0.4 & 0.15 &  0.81 & $f_{\rm bin}=0.5$ (0.3) \\ 
Frag1    & 0.40 &0.34 &0.02 &0.04 &0.00 &  2.4 & 0.28 &    0.9 & 0.25 &  0.63 & $\gamma_{\rm frag}=-0.5$ (0) \\ 
Frag2    & 0.38 &0.28 &0.03 &0.03 &0.00 &  1.1 & 0.09 &   $-$0.5 & 0.08 &  0.67 & $\gamma_{\rm frag}=0.5$ (0) \\ 
Beta1    & 0.39 &0.31 &0.03 &0.04 &0.00 &  1.5 & 0.31 &    0.2 & 0.22 &  0.69 & $\beta=0.5$ (0.7) \\ 
Beta2    & 0.39 &0.30 &0.03 &0.03 &0.00 &  1.2 & 0.06 &    0.5 & 0.04 &  0.74 & $\beta=0.9$ (0.7) \\ 
Fold1    & 0.43 &0.28 &0.06 &0.04 &0.00 &  1.3 & 0.10 &    0.5 & 0.04 &  0.74 & Unfold and shrink \\ 
\hline
\end{tabular}
\end{table*}

\begin{table*}
\center
\caption{Supplementary models for B-type stars}
\label{tab:bpar}
\begin{tabular}{l ccc cc  cc cc c l} 
\hline
Model &  CF & BF & TF & $f_{\rm disrupt}$ & $f_{\rm merge}$ & 
$\gamma_{q,S}$ & $f_{\rm twin,S}$  & $\gamma_{q,L}$ & $f_{\rm twin,L}$  & $\gamma_P$ &
Comment \\
\hline
Observed & \ldots & {\it 0.50} & {\it 0.10} & \ldots & \ldots &  
{\it $-$0.2}  & {\it 0.12}  &  {\it $-$1.5}  & {\it $<$0.05}  &  {\it $-$0.2} &   \\
\hline
Baseline & {\bf 0.75} &{\bf 0.56} & {\bf 0.13} & {\bf 0.45} & {\bf 0.33} &  
{\bf 0.3}  & {\bf 0.09}  &  {\bf $-$1.3}  & {\bf 0.06}  &  {\bf $-$0.40} & Default  \\
PrimSeed1& 0.71 &0.53 &0.12 &0.46 &0.36 &  0.5 & 0.19 &   $-$1.2 & 0.09 & $-$0.35 & $f_{\rm m0}=0.05$ (0.1)  \\  
PrimSeed2& 0.77 &0.56 &0.14 &0.44 &0.32 &  0.4 & $-$0.02 &  $-$1.5 & 0.02 & $-$0.22 & $f_{\rm m0}=0.15$ (0.1) \\ 
CompSeed1& 1.23 &0.73 &0.33 &0.51 &0.08 &  1.4 & 0.12 &   $-$0.6 & 0.10 &  0.14 & $f_{\rm m20}=0.5$ (0.25) \\ 
CompSeed2& 0.13 &0.11 &0.00 &0.15 &0.76 & $-$0.7 & 0.06 &   $-$1.3 & 0.07 & $-$0.74 & $f_{\rm m20}=0.1$ (0.25) \\ 
MaxAcc1  & 1.07 &0.69 &0.25 &0.50 &0.16 &  0.9 & 0.10 &   $-$1.0 & 0.07 & $-$0.17 & $f_{\rm m2max}=2.0$ (1.0) \\ 
MaxAcc2  & 0.26 &0.21 &0.02 &0.26 &0.66 & $-$0.8 & 0.06 &  $-$2.1  & 0.01 & $-$0.58 & $f_{\rm m2max}=0.5$ (1.0) \\ 
Eta1     & 1.31 &0.53 &0.32 &0.42 &0.03 &  1.2 & 0.08 &    0.9 & 0.10 &  0.56 & $\eta=[$-$1,3]$ ([0, 4]) \\     
Eta2     & 0.18 &0.16 &0.01 &0.23 &0.72 & $-$3.5 & 0.00 &    0.0 & 0.00 & $-$0.43 & $\eta=[1,5]$  ([0, 4])   \\ 
Step1    & 1.01 &0.66 &0.22 &0.50 &0.19 &  0.7 & 0.12 &   $-$0.5 & 0.10 & $-$0.16 & $K_{\rm step}=25$ (50)\\ 
Step2    & 0.46 &0.38 &0.05 &0.36 &0.50 & $-$0.3 & 0.08 &   $-$2.5 & 0.00 & $-$0.74 & $K_{\rm step}=100$ (50)\\ 
Mult1    & 0.37 &0.32 &0.03 &0.12 &0.29 &  0.3 & 0.08 &   $-$1.7 & 0.07 & $-$0.50 & $f_{\rm bin}=0.8$ (2.0) \\ 
Mult2    & 0.88 &0.61 &0.18 &0.57 &0.31 &  0.4 & 0.10 &   $-$1.0 & 0.06 & $-$0.42 & $f_{\rm bin}=2.5$ (2.0) \\ 
Frag1    & 0.67 &0.53 &0.11 &0.47 &0.39 &  1.3 & 0.19 &   $-$1.0 & 0.16 & $-$0.53 & $\gamma_{\rm frag}=-0.5$ (0) \\ 
Frag2    & 0.82 &0.57 &0.16 &0.43 &0.29 & $-$0.3 & 0.06 &   $-$1.3 & 0.01 & $-$0.36 & $\gamma_{\rm frag}=0.5$ (0) \\ 
Beta1    & 0.74 &0.55 &0.13 &0.45 &0.34 &  0.2 & 0.18 &   $-$1.6 & 0.15 & $-$0.40 & $\beta=0.5$ (0.7) \\ 
Beta2    & 0.75 &0.56 &0.14 &0.46 &0.33 &  0.6 & $-$0.01 &  $-$1.0 & 0.01 & $-$0.47 & $\beta=0.9$ (0.7) \\ 
Fold1    & 1.46 &0.42 &0.40 &0.49 &0.29 & $-$0.4 & 0.01 &   $-$1.3 & $-$0.02 &  $-$0.11 & Unfold and shrink \\ 
\hline
\end{tabular}
\end{table*}

The  toy  model  is  relatively   robust  to   variation  of  many parameters. For example, model predictions for the companion frequency and twin fraction are relatively insensitive to many parameters except $f_{\rm bin}$. However,  we found that the parameters  that determine the migration rate are critical and affect the results strongly. Those are the migration coefficient $\eta$, the companion seed mass $f_{m20}$, and its growth parameter $f_{\rm m2max}$. Both  small seed companion mass and  limited  growth strongly increase  the migration (and,  for B-type stars,  the merger  rate).   As  a result,  small  mass ratios  become dominant (negative $\gamma_q$). The same effect is produced  by  increasing  the  migration  rate  $\eta$.   The  default combination of these  parameters was chosen by trial  and error to get realistic  outcome  of  the  simulation. Other  combinations  are  not precluded,  but the  strong influence  of these  three  parameters and their  inter-relation is  evident.  The  reason is  that  migration is fastest just after the companion's formation, and its initial mass matters.   When the migration rate increases (larger $\eta$, smaller $f_{m20}$, and/or smaller $f_{\rm m2max}$), the merger rate also increases and thus the close binary fraction drops, which is especially evident for B-type stars.

The following parameters are less  critical.  The initial seed mass of the primary $f_{m0}$ affects mostly  the twin fraction (more twins for smaller seeds). For solar-type binaries,   the number  of accretion  episodes $K_{\rm  step} $  affects the  twin fraction (more  twins if  less episodes), $\gamma_{q,L}$ at long periods (decreases with increasing $K_{\rm step}$), and the slope of  the period distribution (decrease $\gamma_P$), which means stronger  overall migration with  more episodes.  The  effect of changing the companion frequency $f_{\rm bin}$ is obvious: more binaries and  triples,   more  disruptions,  but   almost  no  influence  on  the distributions  of the mass  ratio  and period  for  solar-type stars  (for B-type stars, there is a minor effect). 

Increasing  the  fragmentation   probability  in  the  first  episodes ($\gamma_{\rm frag} = -0.5$) increases the average mass ratio and the twin fraction  because  companions  have  more time  to  grow.   Obviously, $\gamma_{\rm frag} = 0.5$  has the inverse effect and  yields less massive companions.
The prevalence of close solar-type twins strongly suggests accretion rates are large, or at least highly variable, at early times, whereas models in which the mass accretion rate, and therefore probability of disk fragmentation, steadily increase in time are disfavored. Other   parameters  such  as   period  distribution  and multiplicity fraction  remain   almost  unaffected  by   $\gamma_{\rm frag}$.   The parameter  $\beta$ also  affects  mostly the  mass ratio  distribution (more  twins for $\beta  =0.5$,  opposite for $\beta= 0.9$).  

Finally,  replacing disruptions  by unfolding and shrinkage of the inner binary in {\tt Fold1}  has almost  no effect, except the increased fraction of triples. For solar-type binaries, the fraction of disruptions (or unfoldings) is only 0.03.

\section{Summary and discussion}
\label{sec:disc}

Our toy model encompasses the stochastic nature of disk fragmentation (which provide the initial conditions; Section~\ref{IC}), subsequent accretion onto the binary components (mass growth; Section~\ref{BinAcc}), and angular momentum exchanges and losses (migration; Section~\ref{BinMig}). We do not yet have detailed physical models for these various processes that fully and self-consistently incorporate all of the underlying physics, e.g., eccentric orbits, misaligned disks, magnetic fields, variable and chaotic accretion, and outflows and jets. Motivated by observational constraints and hydrodynamic simulations of certain ideal scenarios, our toy model instead utilizes a few simple analytic prescriptions with inherent variability to encapsulate the complexity and stochasticity of disk fragmentation, accretion, and migration. 

Our toy model naturally reproduces 14 observed features of close multiples: (1) a small ratio TF/BF $\approx$ 5\% of compact triples to close binaries for solar-type systems, (2) which increases to TF/BF $\approx$ 20\% for early-B primaries; (3) an increasing solar-type binary period distribution ($\gamma_{\rm P}$ = d$N$/dlog$P$ = 0.7), (4) whereas early-B binaries are skewed toward very short periods ($\gamma_{\rm P}$~=~$\approx$~$-$0.2); (5) the brown dwarf desert, i.e., the paucity of low-mass companions to solar-type primaries at short periods $P$ $\lesssim$ 100 days, (6) but a prevalence of brown dwarf companions at longer periods; (7) the existence of $q$~=~0.05\,-\,0.10 companions to early-B primaries with very short periods $P$~$<$~10~days; (8) a large excess fraction $f_{\rm twin,S}$~$\approx$~0.24 of twins ($q$ $>$ 0.95) among short-period solar-type binaries, (9) which decreases linearly in frequency with log\,$P$; (10) a reduced but non-zero excess twin fraction $f_{\rm twin,S}$ $\approx$ 0.12 among the closest early-B binaries, but (11) no excess twins to early-B primaries beyond $P$~$\gtrsim$~50~days ($f_{\rm twin,L}$~$<$~0.05); (12) a relatively uniform or slightly increasing mass-ratio distribution across $q$~=~0.3\,-\,0.95 both for solar-type binaries ($\gamma_{\rm q}$ $\approx$ 0.5) and (13) for short-period companions to early-B stars ($\gamma_{\rm q,S}$ $\approx$ 0.0), but (14) wider companions to early-B stars that are weighted toward small mass ratios ($\gamma_{\rm q,L}$ $\approx$ $-$1.5).

Twins  correspond to binaries that formed early and experienced a  runaway growth of the mass  ratio.  This process is almost scale-free,  i.e.  independent of  the total mass.   However, a significant  fraction  of  massive  binaries that  formed  early  have merged,  reducing the  fraction  of twins  among  massive stars.  The fraction of twins slowly decreases with increasing period. Early findings that solar-type twins have only short periods $P < 30$\,d \citep{Lucy1979,Tok2000} is likely a selection effect because  double-lined twin spectroscopic binaries require a high spectroscopic resolution to split the lines. However, it is now established that the excess twin population exists among visual binaries \citep{Moe2017,ElBadry2019b}, in line with our predictions.

The brown  dwarf desert is  a natural consequence  of accretion-driven migration.   A  substellar  companion   inevitably  grows   into  the stellar-mass regime while migrating  inward. Only companions formed by disk fragmentation at the very end  of the mass assembly have a chance to  remain substellar \citep{Kratter2010}, but  they migrate little.

Our  model predicts  a large  number  of mergers  during formation  of massive  stars owing  to strong  accretion and,  consequently, fast migration. The latter also  translates into the period distribution that grows toward  short periods and  falls abruptly at the  minimum period corresponding to merger. Early mergers help to build up stellar masses more gradually, compared to a simple accretion. Very massive stars have a short lifetime, and assembling their mass rapidly by accretion implies very high  (probably unrealistic) accretion rates. Formation of massive stars by merging has been suggested several times as a way to alleviate this problem \citep[e.g.][]{Bonnell1998}. However, direct collisions between stars require a very high stellar density if they happen in a cluster, or the existence of many compact and dynamically unstable triple systems. In our scenario, formation of companions, their migration, and merging is driven only by accretion. Massive stars are assembled from gas, but part of this gas is delivered in the form of companions. The lifetime of companions is longer than the lifetime of massive merger products, hence mergers relax the requirement on the accretion rate need to form massive stars. 

It is most remarkable that the toy models for solar-type and B-type stars have  similar  critical parameters that define the migration rate and the mass-ratio distribution ($\eta$,  $f_{m20}$,  $f_{m2max}$, and $\beta$).  The differences in the close-binary statistics of those populations can be explained only by the difference in their mass, using the same prescriptions with similar parameters. 

Accretion-driven migration is inevitably associated with the growth of the mass ratio. Existence of close binaries with small $q$ has always been a challenge to this theory. The toy model addresses this challenge by postulating that companions form during the whole period of mass assembly, not at the same time as the primary star. In young pre-MS eclipsing binaries, the components are not exactly coeval \citep{Stassun2008,Chew2012}.  Moreover, the migration is strongest at the beginning, in the low-$q$ regime. Our crude prescription allows low-mass companions to migrate to short periods before  substantial mass growth. The physics of companion growth and migration in the low-$q$ regime is  complex and unlikely  to be captured correctly by our prescription. However, the prediction  that low-mass companions can rapidly migrate to short separations is one of the results  of our study.

It is instructive to compare our results with those of \citet{Bate2000}. He studied the evolution of seed binaries immersed in co-rotating cores and accreting most of their mass in a deterministic and conservative way according to the prescriptions of  \citet{Bate1997}. In common with this study, Bate predicted  asymptotic growth of the mass ratio towards $q=1$ (formation of twins) and the brown dwarf desert at short periods. However, his model is unable to produce close massive binaries with small $q$, while the initial orbital separations tend to increase, rather than shrink, owing to the conservation of angular momentum. In contrast, our toy model postulates orbit shrinking (negative and random $\eta$), assumes companion formation at random times, rather than simultaneously with the primary, and includes mergers. This helps us to reach a qualitative agreement of the $P,q$ statistics with observations. However, our model hides the complexity of the real binary evolution behind crude prescriptions with random parameters, while Bate based his study on the actual, albeit incomplete, treatment of the binary evolution. We hope that in the future both approaches will converge and that our study will motivate further simulations of accreting binaries leading to better prescriptions for their evolution.


M.M. acknowledges financial support from NASA under Grant No. ATP-170070.

\bibliographystyle{mnras}
\bibliography{migration}

\section{Supplementary material}

The IDL code used to simulate binaries is provided in {\tt toymodel.tar.gz}.

\bsp	
\label{lastpage}
\end{document}